\documentclass[
			reprint,
			amsmath,
			amssymb,
			aps,
			prl,
			]{revtex4-2}

\usepackage{dcolumn}
\usepackage{bm}
\usepackage{graphicx}
\usepackage{xcolor}
\usepackage[latin1]{inputenc}
\usepackage[T1]{fontenc}
\usepackage{amsmath}
	\DeclareMathOperator{\arccosh}{arccosh}
	
	\DeclareMathOperator{\csch}{csch}
	\newcommand{\angstrom}{\text{\normalfont\AA}}
\usepackage{amssymb}
\usepackage{amsfonts}
\usepackage{mathrsfs}
\usepackage{latexsym}
\usepackage{mathdots}
\usepackage{booktabs}
\usepackage{multirow}
\newcolumntype{C}{>{$}c<{$}}
	\AtBeginDocument{
		\heavyrulewidth=.08em
		\lightrulewidth=.05em
		\cmidrulewidth=.03em
		\belowrulesep=.65ex
		\belowbottomsep=0pt
		\aboverulesep=.4ex
		\abovetopsep=0pt
		\cmidrulesep=\doublerulesep
		\cmidrulekern=.5em
		\defaultaddspace=.5em
		}	
\usepackage{hyperref} 
	\hypersetup{  colorlinks,
  citecolor=blue,
  linkcolor=violet,
  urlcolor=blue}
\begin{document}

		\preprint{APS/123-QED}

\title{Cooperative melting in double-stranded peptide chains through local mechanical interactions}
		

\author{Luca Bellino}
	\affiliation{Department of Civil, Environmental, Land and Building Engineering and Chemistry (DICATECh), Polytechnic University Bari, via Orabona 4, 70125 Bari, Italy.}
	\email{luca.bellino@poliba.it}

\author{Giuseppe Florio}
\affiliation{Department of Civil, Environmental, Land and Building Engineering and Chemistry (DICATECh), Polytechnic University Bari, via Orabona 4, 70125 Bari, Italy.}		
\affiliation{INFN, Section of Bari, I-70126, Italy}
		
\author{Alain Goriely}
	\affiliation{Mathematical Institute, University of Oxford, Oxford OX2 6GG, United Kingdom.}
			
\author{Giuseppe Puglisi}
\affiliation{Department of Civil, Environmental, Land and Building Engineering and Chemistry, Polytechnic University Bari, via Orabona 4, 70125 Bari, Italy.}	
	\email{giuseppe.puglisi@poliba.it}
	
%

%
\date{\today}

%
\begin{abstract}

\noindent 
Depending on the external field, double-stranded peptide chains can separate following a cooperative or non-cooperative transition strategy. It is usually believed that these two regimes are driven either by chemical and thermal purple effects, or through non-local mechanical interactions. Here, we show explicitly that local mechanical interactions may regulate the stability and reversibility of the cooperative/non-cooperative debonding transition, which is described by a single parameter depending on an internal length scale. Our theory is applicable to a wide range of fundamental biological examples such as DNA, protein secondary structures, and the microtubules and tau proteins, for which we are able to predict the critical melting force as a function of the chain length.
\end{abstract}
	
%
\maketitle

%

%
\begin{figure}[t]
  \includegraphics[width=1\columnwidth]{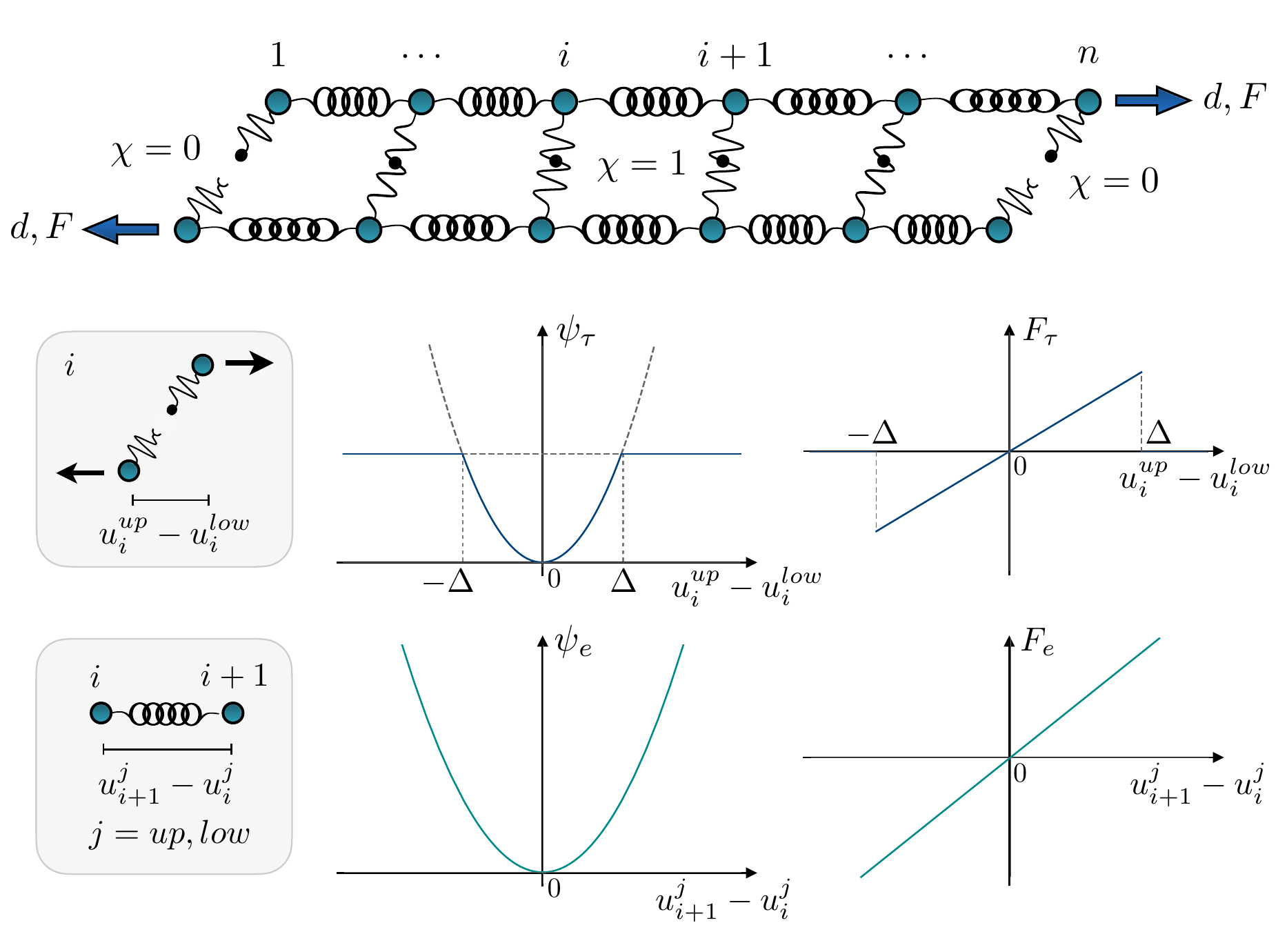}
  \caption{Model describing a generic double-stranded chain with energy and force both of a single breakable crosslink and of the harmonic unit composing the strands.}
   \label{model}
 \end{figure}
%

Polypeptide molecules such as DNA and proteins are the basic structural and information components in living systems. Their assembly and function are of paramount importance to avoid diseases and genetic defects. It is now appreciated that they operate not only through chemical and electrical effects, but also using mechanical and thermal fields~\cite{rief:1999.2, smith:1996, bustamante:2000}. Often, biomolecules undergo conformational transitions to perform tasks such as folding in proteins~\cite{dobson:2003, florio:2021}, DNA replication and denaturation~\cite{watson:1953, bustamante:2000}, axonal growth~\cite{oliveri:2022}, or focal adhesion mechanosensing~\cite{cao:2015, buehler:2007, distefano:2022}. The relative stability of different equilibrium states and their transition reversibility are particularly important for the associated biological functions. 

%
Here, we focus on the {cooperativity}, stability, and toughness of the mechanically induced unfolding transition in double-stranded molecules. \textit{Cooperativity}, the emergent property of a transition based on the coordination of a network of interactions between different segments of the molecule, depends on the external field that drives the transition either through thermal~\cite{lumry:1966} or chemical~\cite{jha:2014} effects. However, mechanical loading may lead to different responses, and variations in linker stiffness can strongly decrease the cooperativity~\cite{hyeon:2005, giordano:2021, shin:2009}. We show that, in a purely mechanical setting, cooperativity and stability effects can be induced by purely mechanical {\it local} interactions. These results are directly relevant for DNA and RNA hairpins~\cite{antao:1991} and to support the hypothesis of a zipper cooperativity model for protein folding~\cite{dill:1993, kloss:2008} considered as a possible solution to Levinthal's paradox.

%
Specifically, we consider a prototypical discrete lattice, modelling two harmonic chains interacting through breakable shear links, suitable for the description of DNA as well for microtubules and tau proteins in axons~\cite{degennes:2001, chakrabarti:2009, prakash:2011}. We obtain analytical solutions capturing crucial features such as the dependence of the rupture force on the length of the molecule, the persistence length, and the transition between cooperative and non-cooperative debonding strategies together with the corresponding transition energy. In particular, for a chain of contour length $L$, we introduce a dimensionless parameter, the \textit{cooperativity index} $\mu=L/\ell_o$, depending on the \textit{localization length} $\ell_o$ that generalizes the notion of persistence length for a single molecule, which regulates the cooperativity and melting energy. We select the value $\mu=1$ ($L=\ell_o$) such that if $\mu\ll 1$ the behavior is an all-or-none phenomenon, whereas if $\mu\gg 1$ the transition is non-cooperative, characterized by a much higher stability of partially debonded states. Interestingly, what emerges from our analysis is that  cooperative transition can be obtained as a purely {\it local} mechanical effect. This completes previous models regulating cooperativity through {\it non-local} interactions~\cite{peyrard:1995, luca:2020,rouzina:2001,hanke:2008}.

\vspace{0.2cm}
\noindent\textbf{Model --} Starting from the classical Peyrard-Bishop (PB) model~\cite{peyrard:1995}, we consider two chains of $n$ massless points connected by elastic springs with energy 
\begin{equation}
\psi_{e}=\frac{1}{2}k_{e}\ell\sum_{j=up,low}\sum_{i=1}^{n-1}\left(\frac{u_{i+1}^j-u_{i}^j}{\ell}\right)^2,
\label{eq:elasticenergy}
\end{equation}
where $\ell$ is the spring length, $k_e$ their stiffness, and the indices $up$ and $low$ indicate upper and low chains, respectively. The chains are connected by local breakable links of stiffness $k_{\tau}$ (see Fig.~\ref{model}), modelling non-covalent interactions such as H-bonds, and subjected to a force $F$ or end-point displacement $d$. Observe that the cartoon in Fig.~\ref{model} schematizes a one-dimensional system with shear springs having negligible reference length and elongation $u^{up}_i-u^{low}_i$ whereas a different behavior is observed in the case of unzipping when bending effects cannot be neglected~\cite{zhang:2016}. Moreover, we assume that shear links have an harmonic regime up to a displacement threshold $\Delta$, followed by a zero force detached state. We notice that, although recrosslinking effects are typical under cyclic loading for biological systems, here we neglect them because we consider monotonical stretching. Thus, by introducing a `spin' variable $\chi_i$ with $\chi_i=1$ ($\chi_i=0$) corresponding to attached (broken) states~\cite{luca:2019, geppe:2013,glauber:1963}, the interaction energy reads 
\begin{equation}
\psi_{\tau}=\frac{1}{2}k_{\tau}\ell\sum_{i=1}^{n}\left[\chi_i\left(\frac{u_i^{up}-u_{i}^{low}}{\Delta}\right)^2+\left(1-\chi_i\right)\right].
\label{eq:shearenergy}
\end{equation}
After introducing the non-dimensional displacements, force, and energy
\begin{equation}
w_i=\frac{u_i}{\Delta},  \quad
\delta=\frac{d}{\Delta}, \quad
f=\frac{F L}{k_e \Delta},\quad
\varphi=\frac{\ell\left(\psi_e+\psi_{\tau}\right)}{k_e \Delta^2}, 
\label{eq:nondimensional}
\end{equation}
we minimize the mechanical energy at assigned displacement $\delta=w_{n}^{up}=-w_1^{low}$, 
\begin{equation}
\min_{w^{up}_i,w^{low}_i}\left[ \varphi(w^{up}_i,w^{low}_i)-\frac{f}{n}\left(w^{up}_n-w^{low}_1-2\delta\right)\right],
\label{eq:varproblem}
\end{equation}
where the force $f$ represents a Lagrange multiplier. We remark that to focus on the role of local mechanical interaction here we neglect external thermal and chemical fields and minimize the total mechanical  (elastic plus fracture) energy $\varphi$ as in classical Griffith type approaches to fracture and decohesion \cite{griffith:1921}.

Equilibrium solutions~\eqref{eq:varproblem} can be parameterized by the configuration $\chi_i$. The exact solution, involving the inversion of tri-diagonal Hessian matrices (see SM), is characterized by shears ($v_i=w_i^{up}-w^{low}_i$) attaining the maximum value at the system boundary~\cite{hu:1996,nabben:1999}. We then obtain that all stable equilibrium solutions are characterized by a single attached segment. Thus, we restrict our attention to `\textit{single-block}' solutions characterized only by the number $p$ of attached bonds, with
\begin{equation}
\begin{split}
\varphi_{eq} &=\kappa^t\delta^2+\mu^2\bigl(1-p/n\bigr),\qquad f=\kappa^t\delta,
\end{split}
\label{eq:energyforce}
\end{equation}
with  $\kappa^{t}\!=4n/(2n\!-\!p\!-\!1\!+\!4\gamma)$ being the chain stiffness where $2\gamma=(\sinh\lambda\!+\!\sinh(p\lambda))/[\sinh((p\!+\!1)\lambda)\!-\!\sinh\lambda\!-\!\sinh(p\lambda)]$ and $\lambda$ is given by (SM18). As previously mentioned, we introduced the non-dimensional \textit{cooperativity index $\mu$}:
\begin{equation} 
\mu^2 =\frac{L^2}{\ell_o^2}=\frac{k_{\tau}}{2  k_e}\frac{L^2}{\Delta^{2}},
\qquad
\ell_o=\sqrt{\frac{2 k_e}{k_{\tau}}}\Delta,
\label{eq:mu}
\end{equation}
where $\ell_o$ is the \textit{localization length} that measures the localization of the decohesion and plays a role similar to the persistence length for single chains~\cite{rubinstein:2003}. These solutions are defined for $\delta\in[0,\delta_{\text{max}}(p)]$ where, based on previously recalled monotonicity results, $\delta_{\text{max}}(p)$, given by (SM32), is the assigned  displacement inducing the crosslink $i=p$ attaining the debonding threshold $u_p^{up}-u_{p}^{low}=\Delta$. 

%

%
\begin{figure*}[t!]
  \includegraphics[width=1\textwidth]{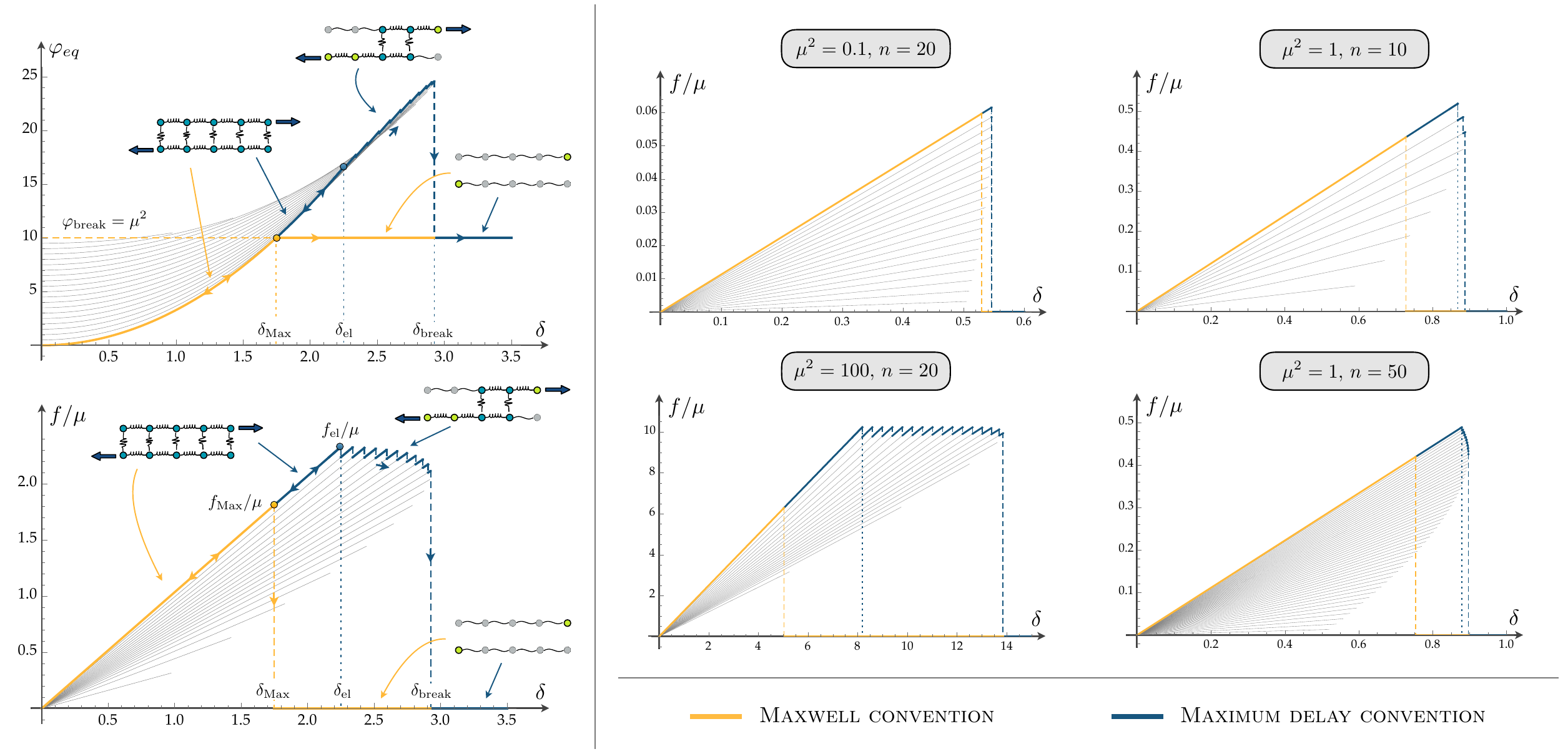}
   \caption{Melting transition behavior in terms of energy and force under the \textit{Maxwell} and the \textit{maximum delay} conventions for a system with $\mu^2=10$, $n=20$. Right: force diagrams for different $\mu^2$ and $n$.}
  \label{fig:meltingstrat}
\end{figure*}

\vspace{0.2cm}
\noindent\textbf{Melting strategies --} The transition of the system in its wiggly energy landscape relies on the competition between loading rate and rate of overcoming energy barriers~\cite{hanggi:1990}. Three main time scales are involved: $\tau_{\,\text{int}}$, the time scale required for exploring the energy wells and relaxing to the local energy minima; $\tau_{\,\text{ext}}$, associated with the external load; $\tau_{\,\text{bar}}$, the time scale to relax to the global energy minima. Here we assume $\tau_{\,\text{int}}\ll \tau_{\,\text{bar}}$ and we consider two limiting cases. First, for high temperature and low rate of loading ($\tau_{\,\text{bar}}\ll\tau_{\,\text{ext}}$), the system configuration always corresponds to the global energy minima (\textit{Maxwell convention}). Second, for  low temperature and rate of loading ($\tau_{\,\text{int}} \ll \tau_{\,\text{ext}} \ll \tau_{\,\text{bar}}$), according with the \textit{maximum delay convention} \cite{geppe:2013} we assume that the system remains in a metastable equilibrium branch (characterized by a certain value of $p$) until it becomes unstable at $\delta=\delta_{\text{max}}(p)$, and these solutions are represented by the gray lines in Fig.~\ref{fig:meltingstrat}.

Under the Maxwell convention (Fig.~\ref{fig:meltingstrat}) the system behaves cooperatively (for all values of $\mu$ and $n$) and follows elastically the fully attached branch $p=n$ until at $\delta_{\text{Max}}$ (yellow path in Fig.~\ref{fig:meltingstrat}) it undergoes a transition to the fully detached state.

\begin{figure*}[ht!]
  \includegraphics[width=1\textwidth]{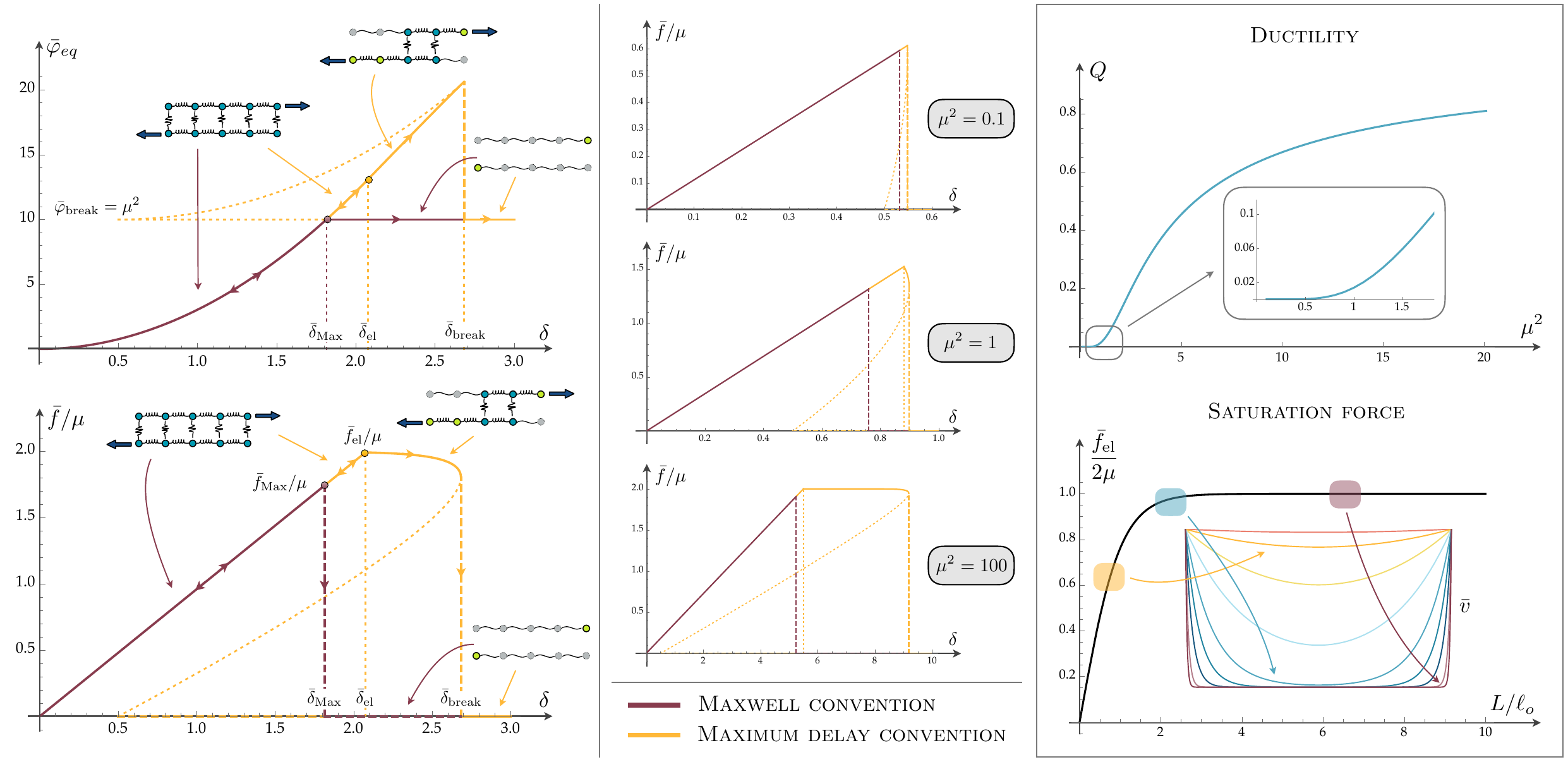}
   \caption{Force and energy in the \textit{continuum limit} for a system with $\mu^2=10$ in the two melting conventions. Centre: force diagrams  varying $\mu^2$. Right: elasto-plastic properties in terms of ductility (measured by the parameter $Q$, see the text for details) and critical force (inset: distribution of the shear vector along the chain).}
  \label{fig:continuum}
\end{figure*}

Under the maximum delay convention (blue lines in Fig. 2) the system switches from a given branch $p=q$ to a branch with $p=q-1$ unbroken bonds when $\delta=\delta_{\text{max}}(q)$. Interestingly, cooperativity is regulated by $\mu$ (see the right panel of Fig.~\ref{fig:meltingstrat}). Decohesion is cooperative for low values of $\mu$ corresponding to high stiffness of the covalent \textit{vs} non-covalent bonds. In this case, after an elastic --reversible-- regime ({$\delta\leq \delta_{\text{el}}$}),  a ductile decohesion is observed, with a sawtooth force-displacement path representing the sequential phase-switching of the domains typically observed in the experiments~\cite{rief:1999}. As the assigned  shear increases, the decohesion front coherently propagates until a second threshold $\delta_{\text{break}}$ is attained and the system fully detaches. Remarkably, as $\mu$ increases ($\ell_o$ decreases)  the ductility of the system $Q=({\delta_{\text{break}} -\delta_{\text{el}}})/{\delta_{\text{el}}}$  increases as its unfolding energy: $\int_0^{\Delta} f/(2\mu)$ (see Fig.~\ref{fig:continuum}, right box). 

We also study the \textit{continuum approximation}, relevant when the number of bonds is large. We fix the contour length $L=n\ell$, consider $n\to\infty$ and $\ell\to0$, and introduce the fraction of unbroken elements $\pi=p/n\in(0,1)$. Using classical arguments~\cite{geppe:2009}, the overall stiffness of the system reads $\bar \kappa^{t}(\pi)=\lim_{n\,\to+\infty}\kappa^{t}=4\mu/[\mu(2-\pi)+\coth(\mu\pi)]$, with force and  energy given by \eqref{eq:energyforce}.

%

Also in this limit under the Maxwell convention the behavior is always cooperative with an all-or-none transition at $\bar\delta_{\text{Max}}\!=\!\sqrt{\mu^2+\mu\coth \mu}/2$ at $\bar f_{\text{Max}}\!=\!{2\mu ^2}/{\sqrt{\mu^2+\mu\coth \mu}}$ (Fig.~\ref{fig:continuum}). On the other hand, within the maximum delay convention, the cooperativity is regulated by $\mu$ and the elastic path holds until the first element breaks at $\bar \delta_{\text{el}}\!=\!(1+\mu\tanh\mu)/2$  at $\bar f_{\text{el}}\!=\!2\mu\tanh \mu$. This force combined with~\eqref{eq:mu} and~\eqref{eq:nondimensional} determines how the critical decohesion force is regulated by the physical  parameters as
\begin{equation}
\bar F_c=\sqrt{2 k_{\tau}k_e}\tanh\left(\sqrt{\frac{k_{\tau}}{2 k_e}}\frac{L}{\Delta}\right).
\label{eq:fcnostra}
\end{equation}
We remark that $\mu$ regulates the maximum force according to~$\bar f_{\text{el}}$ and controls the  cooperativity through the microscopic mechanical properties. Indeed, $\mu$ can be varied either by changing the contour length $L$, or by changing $\ell_o$ which depends on bonds stiffness. This reliance was already pointed out by De Gennes~\cite{degennes:2001}, who questioned the dependence of the rupture force of a dsDNA in terms of the number of base pairs. Conversely, when the length is fixed, we can study the forcex varying bonds properties. The dependence on $\mu$ of the cooperative/non-cooperative transition is crucial also in the limits
\begin{equation}
\begin{split}
\bar{F}_c^{\,\text{linear}} &:=\lim_{\mu\to0}\bar F_c  =2\mu^{2}\frac{k_e \Delta}{L}=\frac{k_{\tau} L}{\Delta},\\
\bar{F}_c^{\,\text{plateau}}&:=\lim_{\mu\to\infty}\bar F_c  =2\mu\frac{k_e \Delta}{L}=\sqrt{2k_{\tau}k_{e}},
\end{split}
\label{eq:limitforces}
\end{equation}
obtaining a linear behavior for small values of $\mu$ and a plateau for large values of $\mu$. 

In the box of Fig.~\ref{fig:continuum} we summarize the influence of $\mu$ on the cooperativity, the stability properties of the melting transition, and the ductility $Q$ with two regimes separated by the threshold  $\ell_o\approx L$. For smaller values of $\mu$ the behavior is fragile. When $\mu$ is large, the behavior is non cooperative, with a ductile melting and a reserve after debonding begins. We give also the saturation force as a function of $L$. For small $\mu$ the shear $\bar{v}$ (see SM) affects the entire chain and the force increases linearly with $L$. As $\mu$ increases, the effect of the shear force is confined to the ends of the chain, leading to a force plateau and a melting force independent from  $L$.

%

%
\begin{figure}[t!]
  \includegraphics[width=1\columnwidth]{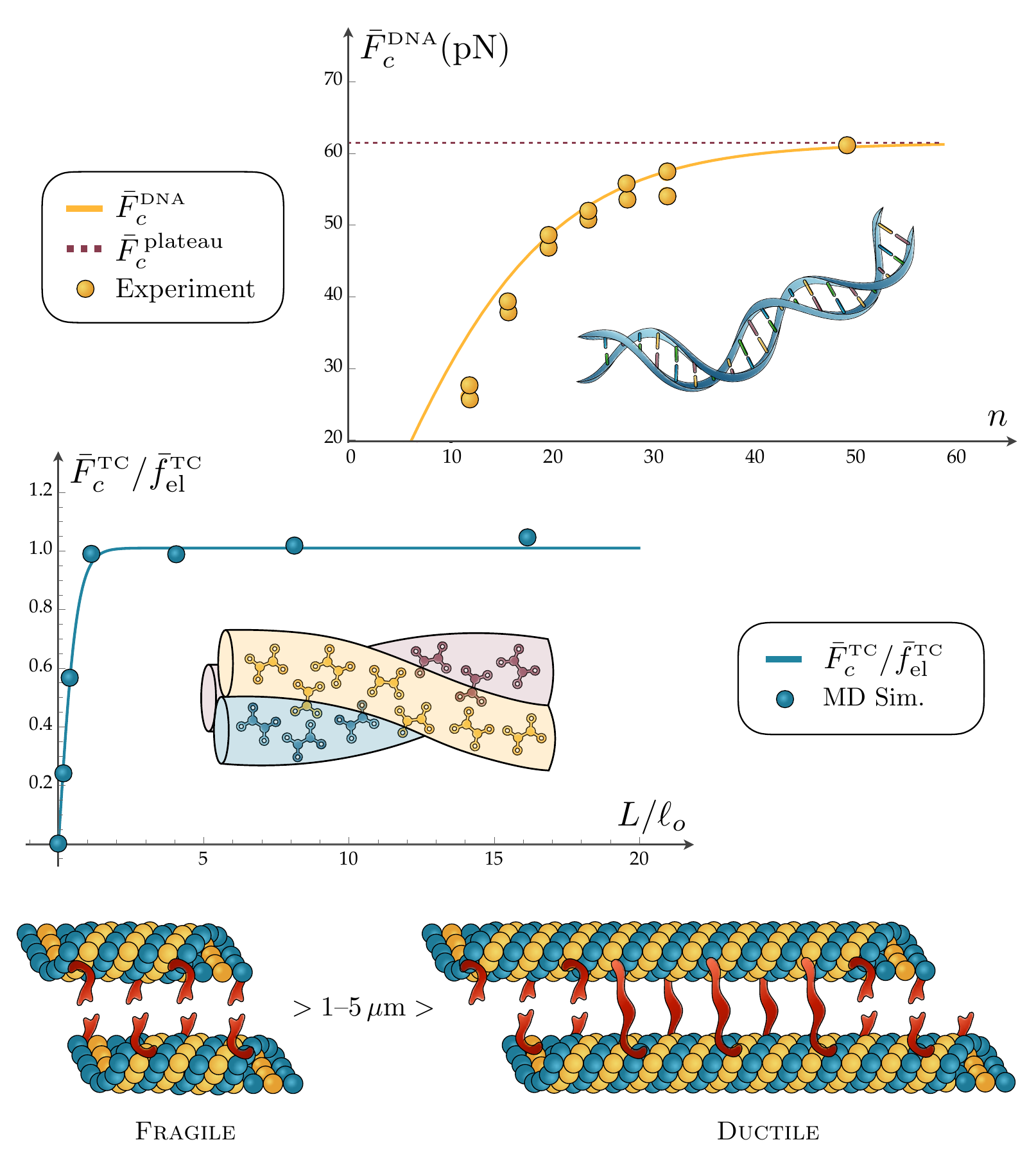}
  \caption{Comparison of our model with various class of experiments on biological systems. See text for parameters.}
\label{fig:experiments}
\end{figure}

\vspace{0.2cm}
\noindent\textbf{Experimental comparison --} Our theoretical results are applicable to diverse biomolecules. 

First, we predict the melting forces of dsDNA with length ranging from $12$ to $50$ base pairs~\cite{hatch:2008}. The theory used in~\cite{hatch:2008} is based on De Gennes's approach~\cite{degennes:2001} but, unfortunately, it requires unphysical values for some  parameters. Here, we use accepted physical parameters~\cite{chakrabarti:2009,gittes:1993, mallik:2004, hirokawa:1988,mosayebi:2015,peyrard:2008,barbi:2003,williams:2002}. Specifically, the distance between bases is $\ell^{\,\text{\tiny{DNA}}}=3.4$ $\angstrom$, $\Delta^{\text{\tiny{DNA}}} \simeq 0.02 - 0.05$ nm ($\Delta^{\text{\tiny{DNA}}}=0.025$ nm in Fig.~\ref{fig:experiments}). To obtain the values of the stiffness, we compared the experimental value of the force plateau $\bar{F}_c^{\,\text{plateau}}=61.5$ pN with the expression of $\bar{F}_c^{\,\text{plateau}}$ in~\eqref{eq:limitforces} obtaining $k_{\tau}^{\,\text{\tiny{DNA}}}k_e^{\,\text{\tiny{DNA}}}\simeq1890$ $\mbox{pN}^2$. Then, by fitting the first three experimental values and comparing them with the expression of the linear force $\bar{F}_c^{\,\text{linear}}$ in~\eqref{eq:limitforces} we obtain $k_{\tau}^{\,\text{\tiny{DNA}}}=0.24$ pN and $k_e^{\,\text{\tiny{DNA}}}=7875$ pN. The obtained values are in perfect agreement also with the ones obtained in~\cite{florio:2023} to fit the thermomechanical melting of DNA. We may then compute also $\mu^{\text{\tiny{DNA}}}\simeq 0.05\,n$ and $\ell_o^{\,\text{\tiny{DNA}}} = 6.5$ nm. Thus, we predict a non-cooperative behavior for $n>n^{\text{\tiny{DNA}}}\simeq 20$ in agreement with~\cite{mosayebi:2015, hatch:2008, chakrabarti:2009}. 

Second, we look at collagen fibrils forming tropocollagen molecules, for which we find the parameters normalized with respect to the elastic force $\bar{f}_{\text{el}}^{\,\text{\tiny{TC}}}$. Using literature data and with the same procedure reported above, we get the rescaled values (see SM) $k_{e}^{\,\text{\tiny{TC}}}=0.3$, $k_{\tau}^{\,\text{\tiny{TC}}}=1.7$ with $\Delta^{\text{\tiny{TC}}}=0.19$ nm and $\ell^{\,\text{\tiny{TC}}}=0.18$ nm~\cite{buehler:2006}. Our theoretical results are in good agreement with the molecular dynamic simulations in Fig.~\ref{fig:experiments}. 

Third, we predict the response of microtubules bounded by tau proteins under shear. According to the experimental literature, we have $k_e^{\,\text{\tiny{MT}}}=1200$ MPa, $k_{\tau}^{\,\text{\tiny{MT}}}=10$ MPa, $\Delta^{\text{\tiny{MT}}}=65$ nm and $\ell^{\,\text{\tiny{MT}}}=30$ nm, obtaining $\mu^{\,\text{\tiny{MT}}}\simeq 0.03\,n$ and $\ell_o^{\text{\,\tiny{MT}}}\simeq 1\,\mu$m, in good agreement with~\cite{ahmadzadeh:2015, ahmadzadeh:2014, meaney:2000, kuhl:2016, hadrien:2022, hawkins:2010}, where it is suggested that  short microtubules ($1\text{--}5\,\mu$m) are fragile but, as length increases, their behavior becomes ductile with a sequential detachment of the tau proteins.

\vspace{0.2cm}
\noindent\textbf{Discussion --} Our theoretical results highlight the fundamental role of local interactions in cooperativity of melting process in biomolecule chains. Under shear-type loading we found different melting regimes. 
In the cooperative (fragile) regime, an all-or-none transition is attained, as in dsDNA~\cite{smith:1996}. Similar effects are found for orthogonal loading~\cite{geppe:2005}, in lattice systems, and in the field of force transmitted by single fibers of composite materials~\cite{marigo:2004, pugno:2013, pugno:2003, distefano:2022}, indicating a saturation force and an internal length scale leading to the so-called `shear lag theory'. 

On the other hand, the non-cooperative --ductile-- melting results in a progressive detachment of the crosslinks, and the `elasto-plastic' properties of such phenomenon are regulated by the cooperativity index (Fig.~\ref{fig:continuum}). In biological systems also thermal and rate effects can have a crucial role with the possibility of spontaneous transition up to a certain critical temperature~\cite{schumakovitch:2002, florio:2022}. Specifically, for low rates thermal fluctuations can drive the system over the energy barrier with unbinding forces smaller than the ones observed in the purely mechanical case, whereas for higher rate the mechanical potential quickly overwhelms the most prominent energy barrier along the reaction pathway, leaving all states free to unbind~\cite{evans:1997, pope:2001, heussinger:2011}. What emerges from our model and analysis is that a non-cooperative $\rightarrow$ cooperative transition  is obtained as an effect of purely local interactions. Other models analyze the important effect of non-local interactions~\cite{peyrard:1995,rouzina:2001,hanke:2008,manghi:2012,cizeau:1997}.

Thus, we have shown that cooperativity, melting stability, and ductility emerge from \textit{local} mechanical interactions in double-stranded chains. With solely one parameter, it is possible to predict different melting regimes crucial in multiple biological systems and in the perspective of designing new molecular-scale devices.

\vspace{+0.5cm}
\paragraph{\textbf{Acknowledgments --}} We thank unknown Referees for useful suggestions and comments that help us improve the clarity of this paper.

%
\vspace{+0.5cm}
\paragraph{\textbf{Fundings --}} LB, GF and GP have been supported by the Italian Ministry MIUR-PRIN project 2017KL4EF3 and by GNFM (INdAM). GP and GF are supported by the project PNRR, National Centre for HPC, Big Data and Quantum Computing (CN00000013) - Spoke 5 ``Environment and Natural Disasters''. GF is supported by INFN project QUANTUM. The work of AG was supported by the Engineering and Physical Sciences Research Council grant EP/R020205/1.

%


%
%

%
%

\newpage

\pagebreak

\clearpage

\widetext

\section{Supplementary Material}
\label{sec:suppmat}
\renewcommand{\theequation}{SM\arabic{equation}}
\setcounter{page}{1}
\renewcommand{\thefigure}{SI\arabic{figure}}
\setcounter{figure}{0}
\setcounter{equation}{0}

\noindent We report  the analytical details and additional information of the model proposed in the paper. In the main manuscript we refer to the equations in this Supplementary Material by denote them with SM($\cdot$) the equation ($\cdot$).

%
\section{Mechanics} 

\subsection{Hamiltonian}
Consider the top (superscript $t$) and bottom (superscript $b$) chains of $n$ massless points, linked by elastic units,with total contour length $L=n\ell$, where $\ell$ is the spring resting length, connected by $n$ breakable crosslinks, as shown in Fig. 1 of the main manuscript. To model the two-phase behavior of the shear units undergoing rupture after a threshold $\Delta$,we introduce the `spin' variable $\chi_i$ assuming values $\chi_i=1$ if the link is attached and $\chi_i=0$ whenbroken. We assume that the distance between the top and bottom chain is negligible, \textit{i.e.} the system can be considered as one-dimensional. Thus, the elongation of the shear springs are $u^{up}_i-u^{low}_i$. The Hamiltonian of the system then reads
\begin{equation}
\psi=\psi_e+\psi_{\tau}=\frac{1}{2}k_{e}\ell\sum_{j=up,low}\sum_{i=1}^{n-1}\left(\frac{u_{i+1}^j-u_{i}^j}{\ell}\right)^2+\frac{1}{2}k_{\tau}\ell\sum_{i=1}^{n}\left[\chi_i\left(\frac{u_i^{up}-u_{i}^{low}}{\Delta}\right)^2+\left(1-\chi_i\right)\right],
\label{eqSM:energy}
\end{equation}
where $k_e$ and $k_{\tau}$ are the stiffnesses of the elastic and shear elements, respectively. In order to obtain non-dimensional quantities we divide the energy by $\ell/(k_e\Delta^2)$, so that we obtain the rescaled displacement $w_i=u_i/\Delta$ and the non-dimensional energy of the system 
\begin{equation}
\varphi(w_i^{up},w_i^{low})=\frac{\ell}{k_e \Delta^2}\psi
=\frac{1}{2}\nu^2\sum_{i=1}^{n}\left[\chi_i\left(w_{i}^{up}-w_{i}^{low}\right)^2+\left(1-\chi_i\right)\right]+\frac{1}{2}\sum_{j=up,low}\sum_{i=1}^{n-1}\left(w_{i+1}^{j}-w_{i}^{j}\right)^2,
\label{eqSM:energynondim}
\end{equation}
where we have introduced the parameter $\nu^2$, defined as
\begin{equation}
\nu^2=\frac{k_{\tau}}{k_e}\frac{\ell^2}{\Delta^{2}}.
\label{eqSM:nu}
\end{equation}
To further simplify the study we introduce the matrices 
\begin{equation}
\boldsymbol{L}=
	\begin{pmatrix}
	2			&-1 		& 		& 		&\boldsymbol{0} 	\\
	-1 			&2		& -1		& 		&				\\
				&\ddots	&\ddots	&\ddots 	&				\\
				&  		& -1		& 2		&-1				\\
	\boldsymbol{0}	& 		&		& -1		& 2				\\
	\end{pmatrix}_{n\times n}, \qquad\qquad
	\boldsymbol{D}=
	\begin{pmatrix}
	\chi_1	&		& 		& 		&\boldsymbol{0} 		\\
			&\ddots	&		&		&		\\
			&		& \chi_i	& 		&		\\																&	 	& 		& \ddots	&		\\
	\boldsymbol{0} 		&		&		& 		&\chi_{n}	\\
	\end{pmatrix}_{n\times n},
	\label{eqSM:matrices}
\end{equation}
where $\boldsymbol{L}$ takes into account the elastic interaction in the upper and lower chains whereas $\boldsymbol{D}$ describes the overall spin states  describing the attached/detached configuration of the breakable links. Moreover,  we introduce the vectors
\begin{equation}
\boldsymbol{w}^{up}=
	\begin{pmatrix}
	w_1^{up}	\\
	w_2^{up}		\\
	\dots	\\
	w_i^{up} 		\\
	\dots	\\
	w_n^{up}		\\ 
	\end{pmatrix}_{n}, \qquad
	\boldsymbol{w}^{low}=
	\begin{pmatrix}
	w_1^{low}		\\
	w_2^{low}		\\
	\dots 	\\
	w_i^{low} 		\\
	\dots	\\
	w_n^{low}		\\ 
	\end{pmatrix}_{n}, \qquad
	\boldsymbol{\chi}=
	\begin{pmatrix}
	\chi_1	\\
	\chi_2	\\
	\dots	\\
	\chi_i	\\
	\dots	\\ 
	\chi_{n}	\\
	\end{pmatrix}_{n}, \qquad
	\boldsymbol{i}_{n}=
	\begin{pmatrix}
	0	\\
	0	\\
	\dots\\
	\dots\\
	0 	\\
	1	\\ 
	\end{pmatrix}_{n}, \qquad
	\boldsymbol{i}_{1}=
	\begin{pmatrix}
	1	\\
	0	\\
	\dots\\
	\dots\\
	0 	\\
	0	\\ 
	\end{pmatrix}_{n}.
	\label{eqSM_vectors}
\end{equation}
Accordingly, the energy of the system is %
\begin{equation}
\varphi(w_i^{up},w_i^{low})=\frac{1}{2}\sum_{j=up,low}\left[\left(\boldsymbol{L}+\nu^2\boldsymbol{D}\right)\boldsymbol{w}^{j}\cdot\boldsymbol{w}^{j}-\left(\boldsymbol{w}^{j}\cdot\boldsymbol{i}_{1}\right)^2-\left(\boldsymbol{w}^{j}\cdot\boldsymbol{i}_{n}\right)^2\right]-\nu^2\boldsymbol{D}\boldsymbol{w}^{up}\cdot\boldsymbol{w}^{low}+\frac{1}{2}\nu^2(n-\boldsymbol{\chi}\cdot\boldsymbol{\chi}).
\label{eqSM:energycompact}
\end{equation}
We also introduce the matrix 
\begin{equation}
\boldsymbol{J}(\boldsymbol{\chi})=\boldsymbol{L}+2\nu^2\boldsymbol{D}(\boldsymbol{\chi})
\label{eqSM:j}
\end{equation}
and the vectors $\boldsymbol{v}$ and $\boldsymbol{z}$ defined by
\begin{equation}
\boldsymbol{v}=\boldsymbol{w}^{up}-\boldsymbol{w}^{low},\qquad\qquad\boldsymbol{z}=\boldsymbol{w}^{up}+\boldsymbol{w}^{low},
\label{eqSM:vz}
\end{equation}
representing the shear displacements and the translation of the $i$-th shear spring center of mass, respectively. Thus, the energy reads 
\begin{equation}
\varphi(\boldsymbol{v},\boldsymbol{z})=\frac{1}{4}\left[\boldsymbol{J}\boldsymbol{v}\cdot\boldsymbol{v}-\left(\boldsymbol{v}\cdot\boldsymbol{i}_{1}\right)^2-\left(\boldsymbol{v}\cdot\boldsymbol{i}_{n}\right)^2\right]+\frac{1}{4}\left[\boldsymbol{L}\boldsymbol{z}\cdot\boldsymbol{z}-\left(\boldsymbol{z}\cdot\boldsymbol{i}_{1}\right)^2-\left(\boldsymbol{z}\cdot\boldsymbol{i}_{n}\right)^2\right]+\frac{1}{2}\nu^2(n-\boldsymbol{\chi}\cdot\boldsymbol{\chi}).
\label{eqSM:energyvz}
\end{equation}
%

\subsection{Equilibrium}

We consider the so-called `hard device hypothesis', where the total end-to-end displacement of the chain is fixed. To do so, we impose the displacements of the last upper and first lower units such that $\delta=w_{n}^t=-w_1^b$ (see Fig. 1 of the main manuscript), and we study the variational problem at fixed configuration $\boldsymbol{\chi}$ by minimizing the function 
\begin{equation}
\min_{w_i^{up}, w_i^{low}}\Biggl[ \varphi(w_i^{up}, w_i^{low})-\frac{f}{n}\left(w_n^{up}-w_1^{low}-2\delta\right)\Biggr] =\min_{\boldsymbol{v}, \boldsymbol{z}}\Biggl[ \varphi(\boldsymbol{v},\boldsymbol{z})-\frac{f}{n}\left(\frac{\boldsymbol{z}+\boldsymbol{v}}{2}\cdot\boldsymbol{i}_{n}-\frac{\boldsymbol{z}-\boldsymbol{v}}{2}\cdot\boldsymbol{i}_{1}-2\delta\right)\Biggr]=\min_{\boldsymbol{v}, \boldsymbol{z}} h(\boldsymbol{v},\boldsymbol{z}), 
\label{eqSM:varproblem}
\end{equation}
where $\varphi$ is given by~\eqref{eqSM:energyvz} and $f$ is the Lagrange multiplier 
\begin{equation}
f:=\frac{F L}{k_e \Delta},
\label{eqSM:forcescaling}
\end{equation}
representing the non dimensional force. Equilibrium conditions are given by   
\begin{equation}
\begin{split}
	&	\frac{\partial h(\boldsymbol{v},\boldsymbol{z})}{\partial \boldsymbol{v}}	= \boldsymbol{J}\boldsymbol{v}-v_{1}\,\boldsymbol{i}_{1}-v_{n}\,\boldsymbol{i}_{n}-\frac{f}{n}\boldsymbol{i}_{n}-\frac{f}{n}\boldsymbol{i}_{1}=\boldsymbol{0},\\
	&	\frac{\partial h(\boldsymbol{v},\boldsymbol{z})}{\partial \boldsymbol{z}}	= \boldsymbol{L}\boldsymbol{z}-z_{1}\,\boldsymbol{i}_{1}-z_{n}\,\boldsymbol{i}_{n}-\frac{f}{n}\boldsymbol{i}_{n}+\frac{f}{n}\boldsymbol{i}_{1}=\boldsymbol{0}.
\end{split}
\label{eqSM:equilibriumeqs}
\end{equation}
A direct inspection of the two matrices show that $\boldsymbol{L}^{-1}_{1,n}=\boldsymbol{L}^{-1}_{n,1}=1-\boldsymbol{L}^{-1}_{n,n}=1-\boldsymbol{L}^{-1}_{1,1}$ and that $\boldsymbol{J}^{-1}_{1,n}=\boldsymbol{J}^{-1}_{n,1}$, so that we have
\begin{equation}
\begin{split}
\boldsymbol{z}	&=\frac{\boldsymbol{L}^{-1}\boldsymbol{i}_{n}-\boldsymbol{L}^{-1}\boldsymbol{i}_{1}}{2\boldsymbol{L}^{-1}_{1,n}}\frac{f}{n},\\
\boldsymbol{v}	&=\frac{\left(1+\boldsymbol{J}^{-1}_{1,n}-\boldsymbol{J}^{-1}_{n,n}\right)\boldsymbol{J}^{-1}\boldsymbol{i}_{1}+\left(1+\boldsymbol{J}^{-1}_{1,n}-\boldsymbol{J}^{-1}_{1,1}\right)\boldsymbol{J}^{-1}\boldsymbol{i}_{n}}{\left(1-\boldsymbol{J}^{-1}_{1,1}\right)\left(1-\boldsymbol{J}^{-1}_{n,n}\right)-\left(\boldsymbol{J}^{-1}_{1,n}\right)^2}\frac{f}{n}.
\end{split}
\label{eqSM:vzequilibrium}
\end{equation}
Observe that $\boldsymbol{L}$ and $\boldsymbol{J}(\boldsymbol{\chi})$ are tridiagonal symmetric matrices. They are always non singular except in the special case when all the shear links are broken ($\chi_i=0, i=1,...,n$) which will be considered separately. By reimposing the kinematic constraints it is possible to evaluate the overall stiffness of the system for a generic $\boldsymbol{\chi}$, indeed we obtain 
\begin{equation}
	\delta =w_{n}^{up}=\frac{z_n+v_n}{2}, \quad -\delta= w_{1}^{low}=\frac{z_1-v_1}{2} \quad \rightarrow \quad \delta=\frac{1}{4}\left(z_n-z_1+v_n+v_1\right).
		\label{eqSM:vzequilibriumdelta}
\end{equation}
Following, by computing the first and last elements of two vectors in~\eqref{eqSM:vzequilibrium} and substituting in~\eqref{eqSM:vzequilibriumdelta}, it is possible to obtain the total stiffness of the system at fixed configuration $\kappa(\boldsymbol{\chi})=f/\delta$, which reads 
\begin{equation}
	\kappa(\boldsymbol{\chi})=4n\left[\frac{1-2\boldsymbol{L}^{-1}_{1,n}}{\boldsymbol{L}^{-1}_{1,n}}+\frac{2\left(\boldsymbol{J}^{-1}_{1,n}\right)^2+2\boldsymbol{J}^{-1}_{1,n}+\boldsymbol{J}^{-1}_{1,1}\left(1-\boldsymbol{J}^{-1}_{n,n}\right)+\boldsymbol{J}^{-1}_{n,n}\left(1-\boldsymbol{J}^{-1}_{1,1}\right)}{\left(1-\boldsymbol{J}^{-1}_{1,1}\right)\left(1-\boldsymbol{J}^{-1}_{n,n}\right)-\left(\boldsymbol{J}^{-1}_{1,n}\right)^2}\right]^{-1}.
	\label{eqSM:ktotvectors}
\end{equation}

To obtain analytical formulas, we need the inverse of the tridiagonal matrix $\boldsymbol{J}$. It can be obtained through recursive expressions but, these expressions are difficult to handle and they hide the physical meaning of the mechanical quantities. Alternatively, if we consider a system with all attached links so that $\boldsymbol{D}(\boldsymbol{\chi})=\boldsymbol{I}$, the inverse can be explicitly evaluated as shown in~\cite{hu:1996SM}. Thus, without loss of generality, we can use this by considering a single block of $p$ connected elements that is joined to the remaining $(n-p)$ elastic units. To prove this result, we demonstrate that for a generic assigned $\delta$ and configuration $p$ the largest component of the shear vector $v_i$ is always attained at the boundaries of the connected block. Consider the inverse of a matrix $\boldsymbol{J}$ with dimension $p\in (0,n)$ corresponding to the block of $p$ attached elements, that is always positive definite and $\boldsymbol{J}^{-1}_{i,j}>\boldsymbol{J}^{-1}_{i,j+1}$, for $j\ge i$~\cite{nabben:1999SM}. In this hypothesis $\boldsymbol{J}$ is doubly symmetric, so that using \eqref{eqSM:vzequilibrium} the required monotonicity of the components of $\boldsymbol{v}$ coincide with that of the vector $\boldsymbol{J}^{-1}\boldsymbol{i}_{1}+ \boldsymbol{J}^{-1}\boldsymbol{i}_{n}$. Then, we need to verify that 
\begin{equation}
\boldsymbol{J}^{-1}_{1,i}+\boldsymbol{J}^{-1}_{n,i}>\boldsymbol{J}^{-1}_{1,i+1}+\boldsymbol{J}^{-1}_{n,i+1} \Leftrightarrow \boldsymbol{J}^{-1}_{1,i}-\boldsymbol{J}^{-1}_{1,i+1}>\boldsymbol{J}^{-1}_{n,i+1}-\boldsymbol{J}^{-1}_{n,i}.
\label{eqSM:dimdiblock}
\end{equation}
Thus, we need to show that  $\boldsymbol{J}^{-1}_{1,i}-\boldsymbol{J}^{-1}_{1,i+1}$ decreases as $i$ grows. To obtain this result, we  use the expression of the inverse matrix in~\cite{hu:1996SM} and extend it by considering $i$ as a continuous variable so that%
\begin{equation}
\frac{\partial(\boldsymbol{J}^{-1}_{1,i}-\boldsymbol{J}^{-1}_{1,i+1})}{\partial i}=\frac{\cosh\left[(p-i)\lambda\right]-\cosh\left[(p+1-i)\lambda\right]}{\sinh\left[(p+1)\lambda\right]} <0,\quad \mbox{with} \quad i<p, \quad \nu^2 \ge 0
\label{eqSM:inversegeneral}
\end{equation}
where 
\begin{equation}
\lambda=\arccosh(1+\nu^2).
\label{eqSM:lambda}
\end{equation}
We conclude that the shears are monotonic in the attached block. 

Thanks to  this result,   we can restrict our attention to a system composed of an attached $p$-dimensional block connected to a nock of ($n-p$) harmonic springs. We remark again that all  configurations with the same $p$ are energetically equivalent so that the system does not recognize the position of the block but only its dimension and that the greatest shears are always at the boundaries, identifying one or two fronts of detachment propagating through the center of the block. Thus, the governing matrices can be inverted and in particular $\boldsymbol{J}^{-1}$ can be obtained from~\cite{hu:1996SM}:
\begin{equation}
\boldsymbol{J}^{-1}_{i,j}=\frac{\cosh\left[(p+1-|j-i|)\lambda\right]-\cosh\left[(p+1-i-j)\lambda\right]}{2\sinh\lambda \sinh(p+1)\lambda},
\label{eqSM:ji}
\end{equation}
and $\lambda$ is given in~\eqref{eqSM:lambda}. We can use this result to express the stiffness of a fully attached system of $p$ units (indicated with the apex $a$) by using~\eqref{eqSM:ji} in~\eqref{eqSM:ktotvectors}: 
\begin{equation}
\kappa^a(p)=\frac{4p}{p-1+4\gamma(p)}, 
\label{eqSM:kattached}
\end{equation}
where
\begin{equation}
\gamma(p)=\frac{\sinh(\lambda)+\sinh(p\lambda)}{2\left\{\sinh[\left(p+1\right)\lambda]-\sinh(\lambda)-\sinh(p\lambda)\right\}}.
\label{eqSM:gammap}
\end{equation}
Also, we compute explicitly the displacement vectors 
\begin{equation}
\begin{split}
z_i & =\frac{2\left(2i-p-1\right)}{p-1+4\gamma(p)}\delta,\qquad i=1,\dots,p\\
&\\
v_i & =\frac{2\left\{\cosh\left[(p-i+2)\lambda\right]-\cosh\left[(p-i)\lambda\right]+\cosh\left[(i+1)\lambda\right]-\cosh\left[(i-1)\lambda\right]\right\}}{\sinh(\lambda)\left\{(p-i)\sinh\left[(p+1)\lambda\right]-(p-3)\left[\sinh(\lambda)+\sinh(p\lambda)\right]\right\}}\delta\qquad i=1,\dots,p.
\end{split}
\label{eqSM:vzexplicit}
\end{equation}
To join the attached and detached parts of the system we consider the kinematic constraints that fixes the total displacement as the sum of the attached and detached elements. First ,consider the detached part, where the displacement respects the equilibrium condition 
\begin{equation}
w^j_{i+1}-w_{i}^{j}=\frac{f}{2n} , \qquad\qquad i=p+1,\dots,n, \qquad j=up,low, 
\label{eqSM:detachedsol}
\end{equation}
where, again, we relabelled the indexes to take care of the arbitrariness of the position of the detached elements. Thus we obtain
\begin{equation}
\delta=\delta^{a}+\delta^{d}=\frac{p}{\kappa^a(p)}\frac{f}{n}+\frac{n-p}{2}\frac{f}{n}.
\label{eqSM:deltatot}
\end{equation}
Recalling that $\kappa^{t}(n,p)=f/\delta$, we obtain the total (superscript $t$) stiffness of the generic $n$-dimensional system at given configuration $p$ as
\begin{equation}
\kappa^{t}(n,p)=n\left(\frac{p}{\kappa^a(p)}+\frac{n-p}{2}\right)^{-1}=\frac{4n}{2n-p-1+4\gamma(p)}.
\label{eqSM:ktotexplicit}
\end{equation}
Therefore, the total energy of the system is 
\begin{equation}
\varphi_{eq}(n,p,\delta)=\kappa^t(n,p)\delta^2+\mu^2\left(1-\frac{p}{n}\right),
\label{eqSM:energyexplicit}
\end{equation}
with a corresponding force
\begin{equation}
f(n,p,\delta)=\kappa^t(n,p)\delta,
\label{eqSM:forceexplicit}
\end{equation}
where we have introduced the \textit{cooperativity index}
\begin{equation}
\mu^2:=\frac{L^2}{\ell_o^2}=\frac{\nu^2n^2}{2}=\frac{k_{\tau}}{2k_e}\frac{L^2}{\Delta^2}, 
\label{eqSM:mu}
\end{equation}
with 
\begin{equation}
\ell_o=\sqrt{\frac{2k_{e}}{k_{\tau}}}\Delta, 
\label{eqSM:lo}
\end{equation}
the \textit{localization length}, that is introduced and discussedin the main manuscript. 

Finally, by using~\eqref{eqSM:vzequilibrium}, the existing domain of each equilibrium branch can be computed by considering the limit condition attained when the external link reach the threshold $v_p=1$:
\begin{equation}
v_p=\frac{1}{\kappa_{\text{break}}(p)}\frac{f}{n}=\frac{\kappa^{t}(n,p)}{\kappa_b(p)}\frac{\delta_{\text{max}}}{n}=1,
\label{eqSM:vlastp}
\end{equation}
where we introduced the stiffness factor $\kappa_{break}(p)$ relating the displacement of the end point elements of the attached block to the force, that can be evaluated by the simple relation
\begin{equation}
\kappa_{\text{break}}(p)=\frac{\gamma(p)}{2}.
\label{eqSM:kbrokenp}
\end{equation}
Thus for the $p$ branch we have
\begin{equation}
\delta \in(0,\delta_{\text{max}}), \quad \delta_{\text{max}}:=\frac{1}{4} \left\{\frac{(2 n-p-1) \sinh \left[(p+1) \lambda\right]}{\sinh(\lambda)+\sinh(p \lambda)}-2 n+p+3\right\}, \quad p=1,...,n.
\label{eqSM:deltamaxp}
\end{equation}

\subsection{Decohesion strategies}

We show that in the Maxwell convention the behavior is always \textit{fragile}. We prove that the energy branch representing the fully attached configuration (\textit{i.e.} $\varphi_{eq}(n,n,\delta)$) intersects the fracture energy line (fully detached, \textit{i.e.} $\varphi_{eq}(n,0,\delta)$) before intersecting the energy diagrams of all other partially detached solutions. Accordingly, at fixed $n$ and $\nu^2$ we evaluate the displacement $\delta_p$ for each intersection  by solving the  equality: 
\begin{equation}
\varphi_{eq}(p)=\varphi_{eq}(p+1) \Leftrightarrow \delta_p=\frac{\mu}{\sqrt{2n\left[\kappa^t(p+1)-\kappa^t(p)\right]}}.
\label{eq:ch5_deltap}
\end{equation}
To prove our result we need to show  that~\eqref{eq:ch5_deltap} decreases as $p$ decreases. Thus we need to show that 
\begin{equation}
\frac{\partial \delta_p}{\partial p}=\frac{\mu}{2\sqrt{2n}}\frac{\frac{\partial \kappa^t(p)}{\partial p}-\frac{\partial \kappa^t(p+1)}{\partial p}}{\Big[\kappa^t(p+1)-\kappa^t(p)\Big]^{\frac{3}{2}}}>0\quad \forall \,p <n
\end{equation}
which reduces to 
\begin{equation}
\frac{\partial \kappa^t(p)}{\partial p}-\frac{\partial \kappa^t(p+1)}{\partial p}>0\quad \forall \,p <n,
\end{equation}
By a direct inspection of the derivative of~\eqref{eqSM:ktotexplicit}, one observes that the condition is always fulfilled, which proves our statement.

%
\subsection{Continuum approximation} 

Following~\cite{geppe:2005}, we fix the total length of the chain $L=n\ell =\text{const}$ with $n\to\infty$ and $\ell\to 0$. We introduce the fraction of unbidden elements $\pi$, described in this limit as a continuum variable and assuming values in the interval $(0,1)$, and we have
\begin{equation}
\pi:=\frac{p}{n}, \qquad \qquad \iota=\frac{i}{n}
\label{eqSM:rescaledquantities}
\end{equation}
so that the correct scaling of the energy in~\eqref{eqSM:energynondim} is obtained for growing $n$ at fixed $\mu$. From now on, we indicate the relevant quantities by an overbar and we find the total stiffness of the chain is
\begin{equation}
\bar \kappa^{t}(\pi)=\lim_{n\,\to+\infty}\kappa^{t}(n,p)=\frac{4\mu}{\mu(2-\pi)+\coth\left(\mu\pi\right)}.
\label{eqSM:ktot_lc}
\end{equation}
Accordingly, the force-displacement relation and the energy are given by 
\begin{equation}
\begin{split}
\bar f(\pi,\delta)&=\bar \kappa^{t}(\pi)\delta,\\
\bar \varphi(\pi,\delta)&=\bar \kappa^t(\pi)\delta^2+\mu^2\left(1-\pi\right).
\end{split}
\label{eqSM:forceenergy_lc}
\end{equation}
We also compute the shear displacements
\begin{equation}
\bar{v}_{\iota}=\frac{\left\{\csch^2\left(\mu\pi\right) \left[\sinh \left(2 \iota\mu\right)-\sinh\left(2\mu\left(\iota -\pi \right)\right)\right]\right\}}{\pi \left[\mu\left(2-\pi\right)+\coth\left(\mu\pi\right)\right]}\delta
\end{equation}
and
\begin{equation}
\bar{z}_{\iota}=\frac{2\mu\left(2\iota -\pi\right)}{\pi\left[\mu\left(2-\pi\right)+\coth\left(\mu\pi\right)\right]}\delta
\end{equation}
Similarly, by using~\eqref{eqSM:deltamaxp}, we obtain the expression in the continuum approximation of the maximum displacement at given $\pi$:
\begin{equation}
\bar \delta_{\text{max}}(\pi)=\lim_{n\,\to+\infty}\delta_{\text{max}}(n,p)=\frac{1}{2} \Bigl[1+\mu (2-\pi) \tanh \left(\mu\pi\right)\Bigr].
\label{eqSM:deltamax_lc}
\end{equation}

Next, we study  two hypotheses  for the melting strategies in the continuum limit. In the Maxwell convention we write separately the energy and the force for the fully attached system and for the fully detached one and we prove  that even in this case the fragile rupture behavior is always attained. We indicate the fully attached system (resp. fully detached) by the subscripts \textit{fa} (resp. \textit{fd}). Imposing  $\pi=1$ in~\eqref{eqSM:forceenergy_lc}, we obtain 
\begin{equation}
\bar \varphi_{\textit{fa}}(\delta)=\frac{4\mu}{\mu+\coth\mu}\delta ^2 , \quad\quad \bar f_{\textit{fa}}(\delta)=\frac{4\mu}{\mu+\coth\mu}\delta.
\label{eqSM:fasolutions_lc}
\end{equation}
whereas if $\pi=0$, we obtain
\begin{equation}
\bar \varphi_{\textit{fd}}=\mu^2, \quad\quad \bar f_{\textit{fd}}=0.
\label{eqSM:fdsolutions_lc}
\end{equation}
Then, we compute the threshold value for fracture  $\bar{\delta}_{\text{Max}}$ in the Maxwell hypothesis, obtained by imposing $\bar \varphi_{\textit{fa}}(\delta_{\text{Max}})=\bar \varphi_{\textit{fd}}$ and the corresponding Maxwell force as
\begin{equation}
\bar \delta_{\text{Max}}=\frac{1}{2}\sqrt{\mu^2+\mu\coth \mu},\qquad\qquad\bar f_{\text{Max}}=\frac{2\mu ^2}{\sqrt{\mu^2+\mu\coth \mu}}.
\label{eqSM:deltamax}
\end{equation}

Consider now the maximum delay convention, where the system is allowed to stay in a local minimum until that configuration is locally unstable. The fully attached solutions holds until the first breakable element detaches at the `elastic' displacement $\bar{\delta}_{\text{el}}$ with a conjugate force  
\begin{equation}
\bar \delta_{\text{el}}:=\bar \delta_{\text{max}}(1)=\frac{1}{2} \Bigl[1+\mu\tanh\mu\Bigr], \qquad \qquad \bar f_{\text{el}}:=\bar f_{\textit{fa}}(\bar \delta_{\text{el}})=2\mu\tanh\mu.
\label{eqSM:deltael}
\end{equation}
We also show that  in this limit the behavior of the system in the Maxwell convention is always fragile, by noting that the conditions
\begin{equation}
\bar \varphi_{\textit{fa}}(\bar \delta_{\text{el}})>\bar \varphi_{\textit{fd}}, \qquad\qquad\bar \delta_{\text{el}}>\bar \delta_{\text{Max}},
\label{eqSM:energycondition}
\end{equation}
are fulfilled. Indeed, we have
\begin{align}
\bar \varphi_{\textit{fa}}(\bar \delta_{\text{el}})-\bar \varphi_{\textit{fd}}&=\mu^2\tanh^2\mu+\mu\tanh\mu-\mu^2>0 &\text{for}\quad\mu>0,\\
\bar \delta_{\text{el}}-\bar \delta_{\text{Max}}&=\frac{1}{2} \Bigl[1+\mu\tanh \mu\Bigr]-\frac{1}{2}\sqrt{\mu^2+\mu\coth \mu}>0 &\text{for}\quad \mu>0.
\label{eqSM:dim2}
\end{align}
%

\subsection{Physical parameters and $\mu$ limits}

To uncover the physical meaning of the elastic force in~\eqref{eqSM:deltael}, which is the critical force $\bar{F}_c$ at which the system starts the melting process, we use~\eqref{eqSM:mu} and~\eqref{eqSM:forcescaling} to obtain 
\begin{equation}
\bar{F}_c=\sqrt{2 k_{\tau} k_e}\tanh\left(\sqrt{\frac{k_{\tau}}{2k_e}}\frac{L}{\Delta}\right). 
\label{eqSM:fc}
\end{equation}

Consider now two limit cases. First, when the cooperativity index is small (\textit{i.e.} $\mu\to0$), the contour length is much smaller than the localization length $L\ll\ell_o$, and from~\eqref{eqSM:deltamax} we obtain the Maxwell force and displacement 
%
%
%
\begin{equation}
\bar{f}_{\text{Max}}^{\,\mu\to0}=\lim_{\mu\to0}\bar{f}_{\text{Max}}\simeq 2\mu^2+\mathcal{O}(\mu^4), \qquad\qquad \bar{\delta}_{\text{Max}}^{\,\mu\to0}=\lim_{\mu\to0}\bar{\delta}_{\text{Max}}\simeq \frac{1}{2}+\frac{\mu^2}{3}+\mathcal{O}(\mu^4),
\end{equation}
as well as the elastic force and displacement in this regime: 
\begin{equation}
\bar{f}_{\text{el}}^{\,\mu\to0}=\lim_{\mu\to0}\bar{f}_{\text{el}}\simeq 2\mu^2+\mathcal{O}(\mu^4), \qquad\qquad \bar{\delta}_{\text{el}}^{\,\mu\to0}=\lim_{\mu\to0}\bar{\delta}_{\text{el}}\simeq \frac{1}{2}+\frac{\mu^2}{2}+\mathcal{O}(\mu^4).
\end{equation}
In this situation we compute the value of the critical force~\eqref{eqSM:fc} by considering the physical quantities in~\eqref{eqSM:mu} and~\eqref{eqSM:forcescaling} which reads 
\begin{equation}
\bar{F}_c^{\,\text{linear}}:=\lim_{\mu\to0} \bar{F}_c=\bar{f}_{\text{Max}}^{\,\mu\to0}=\bar{f}_{\text{el}}^{\,\mu\to0}=2\mu^2\frac{k_e\Delta}{L}=\frac{k_{\tau}L}{\Delta}.
\label{eqSM:flinear}
\end{equation}
that increases linearly with the length of the system (see the critical force of Fig. 3 in the main manuscript). 

On the other hand, when $L\gg\ell_o$ the melting transition is non-cooperative, thus we study the limit $\mu\to\infty$  and we obtain, as in the previous case, the Maxwell force and displacement 
\begin{equation}
\bar{f}_{\text{Max}}^{\,\mu\to\infty}=\lim_{\mu\to\infty}\bar{f}_{\text{Max}}\simeq 2\mu+\mathcal{O}(\mu^4), \qquad\qquad \bar{\delta}_{\text{Max}}^{\,\mu\to\infty}=\lim_{\mu\to\infty}\bar{\delta}_{\text{Max}}\simeq \frac{\mu}{2}+\mathcal{O}(\mu^4),
\end{equation}
and the same quantities in the maximum delay convention 
\begin{equation}
\bar{f}_{\text{el}}^{\,\mu\to\infty}=\lim_{\mu\to\infty}\bar{f}_{\text{el}}\simeq 2\mu+\mathcal{O}(\mu^4), \qquad\qquad \bar{\delta}_{\text{el}}^{\,\mu\to\infty}=\lim_{\mu\to\infty}\bar{\delta}_{\text{el}}\simeq \frac{\mu}{2}+\mathcal{O}(\mu^4),
\end{equation}
In this case we eventually obtain the value of the force plateau:
\begin{equation}
\bar{F}_c^{\,\text{plateau}}:=\lim_{\mu\to\infty} \bar{F}_c=\bar{f}_{\text{Max}}^{\,\mu\to\infty}=\bar{f}_{\text{el}}^{\,\mu\to\infty}=2\mu\frac{k_e\Delta}{L}=\sqrt{2k_{\tau}k_e}.
\label{eqSM:fplateau}
\end{equation}
%

\section{Experiments}

Next, we give some details on the comparison between our theoretical results and the different experimental data presented in the main manuscript. 

\vspace{0.2cm}
\noindent\textbf{\textit{DNA}} -- The saturation of the critical force~\eqref{eqSM:fc} to a plateau for long chains is a known result, initially developed  in a  continuum framework~\cite{degennes:2001SM}. This result has been used by Hatch and coworkers to fit their experiments performed with magnetic tweezers on either $3'3'$ and $5'5'$ ends of DNA sequences of different length, ranging from $12$ to $50$ base pairs~\cite{hatch:2008SM}. Unfortunately, their analysis requires setting parameters to unphysical values and a modification of the original formulation by De Gennes. In particular, it leads to the conclusion that stretching hydrogen bonds between base pairs is almost 2 orders of magnitude easier than extending the distance between stacked bases in a duplex~\cite{mosayebi:2015SM}. 

From a fit to the experimental data, we obtain the values of the physical parameters appearing in our theoretical formulation. In particular, by comparing the asymptotic experimental critical force of $\bar{F}_c=61.5$ pN with the formula in~\eqref{eqSM:fplateau} we obtain 
\begin{equation}
k_{\tau}^{\,\text{\tiny{DNA}}}k_e^{\,\text{\tiny{DNA}}}\simeq1.89\times 10^{-21}.
\label{eqSM:prodk}
\end{equation}
We used the first three experimental points corresponding to the linear regime (see Fig.5 of the main manuscript) given by of~\eqref{eqSM:flinear} to obtain the value of $k_{\tau}^{\,\text{\tiny{DNA}}}=0.24$ pN such that, using~\eqref{eqSM:prodk}, the strand stiffness reads $k_{e}^{\,\text{\tiny{DNA}}}=7875$ pN. The distance of the base pairs is a known parameter, $\ell^{\,\text{\tiny{DNA}}}=0.34$ nm whereas the breaking displacement ranges between $\Delta^{\text{\tiny{DNA}}}=0.02-0.05$ nm, so that we choose $\Delta^{\,\text{\tiny{DNA}}}=0.025$ nm. Accordingly, we obtain the cooperativity index and the localization length for the DNA
\begin{equation}
\mu^{\text{\tiny{DNA}}}(n)\simeq0.05\, n, \qquad\qquad \ell_o^{\,\text{\tiny{DNA}}}\simeq6.5\text{ nm}.
\end{equation}
Thus, we obtain a threshold separating a cooperative/non-cooperative behavior of $n^{\text{\tiny{DNA}}}\simeq20$. In Tab.~\ref{tabSM:DNA} we summarize the values used for the DNA. 
\begin{table}[h!]
\footnotesize
\centering
\begin{tabular}{cll}
\toprule
Synbol & Value & Unit\\
\midrule
$k_{e}^{\,\text{\tiny{DNA}}}$		& 7875	& pN \\
$k_{\tau}^{\,\text{\tiny{DNA}}}$	& 0.24	& pN\\
$\Delta^{\text{\tiny{DNA}}}$		& 0.025	& nm \\
$\ell^{\,\text{\tiny{DNA}}}$		& 0.34	& nm\\
\bottomrule
\end{tabular}
\caption{Parameter of DNA.}
\label{tabSM:DNA}
\end{table}
\vspace{0.2cm}
\noindent\textbf{\textit{Tropocollagen fibers}} -- Here we compare our results with the molecular dynamic simulations performed in~\cite{buehler:2006SM} on tropocollagen fibers. With the same reasoning as in the previous case, we compute the value of the force plateau with respect to the molecular dynamic value and normalized it with respect to the elastic force $\bar{f}_{\text{el}}^{\text{\tiny{TC}}}$. We  obtain a normalized value of $k_{e}^{\,\text{\tiny{TC}}}\times k_{\tau}^{\,\text{\tiny{TC}}}=0.5$. Then, by using~\eqref{eqSM:flinear} we obtain $k_{\tau}^{\,\text{\tiny{TC}}}=1.7$ and thus $k_{e}^{\,\text{\tiny{TC}}}=0.3$. The remaining parameters (length and threshold displacement) are taken from~\cite{buehler:2006SM}. In Tab.~\ref{tabSM:mt} we summarize the parameters used to predict the behavior of the saturation force for Tropocollagen fibers.
\begin{table}[h!]
\footnotesize
\centering
\begin{tabular}{cll}
\toprule
Symbol & Value & Unit\\
\midrule
$k_{e}^{\,\text{\tiny{TC}}}$ 		& 0.3		& / \\
$k_{\tau}^{\,\text{\tiny{TC}}}$		& 1.7		& /\\
$\Delta^{\text{\tiny{TC}}}$		& 0.19	& nm \\
$\ell^{\,\text{\tiny{TC}}}$			& 0.18	& nm\\
\bottomrule
\end{tabular}
\caption{Parameter of Tropocollagen.}
\label{tabSM:mt}
\end{table}
%

\vspace{0.2cm}
\noindent\textbf{\textit{Microtubules and tau proteins}} -- To describe the fragile/ductile decohesion process attained by the bundles of microtubules and tau proteins of different lengths, we use  parameters from the literature given in Tab.~\ref{tabSM:mt}. Moreover, we provide a full set of data collected within the development of this work concerning axons, microtubules and tau proteins in Tab.~\ref{tabSM:mtparameters}. Specifically we compute both $\mu^2(n)$ and $\ell_o$ obtaining 
\begin{equation}
\mu^{\text{\tiny{MT}}}(n)\simeq 0.03\, n, \qquad\qquad \ell_o^{\,\text{\tiny{MT}}}\simeq 1\text{ $\mu$m}.
\end{equation}
As described in the main text, this result is in good agreement with other studies, wherea brittle detachment of  tau proteins is observed  for short microbutules ($1-5$ $\mu$m) and a sequential detachment of the cross-linking tau is attained for long MTs ($>5$ $\mu$m).
\begin{table}[h!]
\footnotesize
\centering
\begin{tabular}{cllc}
\toprule
Symbol & Value & Unit\\
\midrule
$k_{e}^{\,\text{\tiny{MT}}}$ 		& 1200	& MPa 	&\cite{gittes:1993SM}\\
$k_{\tau}^{\,\text{\tiny{MT}}}$		& 10		& MPa	&\cite{mallik:2004SM}\\
$\Delta^{\text{\tiny{MT}}}$		& 65		& nm 	&\cite{hirokawa:1988SM}\\
$\ell^{\,\text{\tiny{MT}}}$		& 30		& nm	&\\
\bottomrule
\end{tabular}
\caption{Parameter of Microtubules and tau proteins.}
\label{tabSM:mt}
\end{table}
%

%
\begin{table}[h!]
\footnotesize
\centering
	\begin{tabular}{lllc}
      	\toprule
      								&\textsc{Value}		& \textsc{Unit} 		&\textsc{References}				\\
	\midrule
	\textsc{\textbf{Axon}	}			&				&				&							\\
	\midrule
      	Axon lenght					& 40				& $\mu$m		& \cite{caminiti:2013SM}			\\
	Axon diameter					& 540		 	& nm			& \cite{hirokawa:1982SM}			\\
	MTs per cross section				& 9.5			 	& -				& \cite{bray:1981SM}				\\		Cytosol viscosity				& 5				& mPa s			& \cite{haak:1976SM}			\\
	\midrule
	\textsc{\textbf{Microtubules}}		&				&				&							\\
	\midrule
      	Microtubule lenght				& 2-10			& $\mu$m		& \cite{yu:1994SM}				\\
	Microtubule outer diameter		& 25			 	& nm			& \cite{suresh:2007SM}			\\
	Microtubule inner diameter		& 14			 	& nm			& -							\\
	Microtubule area				& 400			& nm$^2$			& \cite{suresh:2007SM}			\\
	Microtubule stiffness				& 1200		 	& MPa			& \cite{gittes:1993SM}					\\
	Microtubule Young's modulus		& 1.5 - 1.9			& GPa			& \cite{suresh:2007SM,pampaloni:2006SM}	\\
	Microtubule flexural rigidity		& $1.8\times10^{-24}$& Nm$^2$		& \cite{pampaloni:2006SM}		\\
	Microtubule spacings				& 23-38			& nm			& \cite{rosenberg:2007SM}			\\
	Microtubule persistence lenght		& 420			& $\mu$m		& \cite{pampaloni:2006SM}		\\
	Microtubule tensile strain			& 50\%			& -				& \cite{janmey:1991SM}			\\
	Polymerization rate				& 1			 	& nm/ms			& \cite{bunker:2004SM}			\\
	Depolymerization rate			& 2			 	& nm/ms			& \cite{bunker:2004SM}			\\
	Polymerization time				& 2000			& ms				& \cite{bunker:2004SM}			\\
	Depolymerization time			& 1000		 	& ms				& \cite{bunker:2004SM}			\\
	\midrule
	\textsc{\textbf{$\tau$-Proteins}}	&				&				&							\\
	\midrule
      	 $\tau$ protein lenght			& 45			 	& nm			& \cite{mofrad:2012SM}				\\
	 $\tau$ protein area				& 1			 	& nm$^2$			& \cite{kuhl2016SM}				\\
	 $\tau$ protein stiffness			& 10				 & MPa			& \cite{mallik:2004SM}			\\
	 $\tau$ protein Young's modulus	& 5			 	& MPa			& \cite{mofrad:2012SM}			\\
	 $\tau$ protein spring constant		& 25			 	& pN/nm			&\cite{wegmann:2011SM}			\\
	 $\tau$ protein max stretch			& 150 \%		 	& -				&-							\\
	 $\tau$ protein distance			& 1				& nm			& \cite{hirokawa:1982SM}			\\
	 $\tau$ protein spacings			& 20-40			& nm			& \cite{hirokawa:1988SM}			\\
	 $\tau$ protein angle				& 10-45			 & deg			& \cite{hirokawa:1982SM}			\\
	 $\tau$ protein bond force			& 10			 	& pN			& estimated					\\
	 $\tau$ attachment rate			& 4			 	& 1/s			& \cite{wegmann:2011SM}			\\
	  $\tau$ protein dashpot timescale	& 0.35			& s				&\cite{wegmann:2011SM}			\\
      	\bottomrule
    	\end{tabular}
\caption{Properties of axons, microtubules and tau proteins.}
\label{tabSM:mtparameters}
\end{table}
%

%


\end{document}